\documentclass[aps,twocolumn,showpacs,superscriptaddress,floatfix,letter]{revtex4}
\usepackage{graphicx}
\usepackage{psfrag}
\usepackage{amsmath}
\begin{document}
\newcommand{\beq}{\begin{equation}}
\newcommand{\eeq}{\end{equation}}
\newcommand{\ben}{\begin{eqnarray}}
\newcommand{\een}{\end{eqnarray}}
\newcommand{\bea}{\begin{array}}
\newcommand{\eea}{\end{array}}
\newcommand{\om}{(\omega )}
\newcommand{\bef}{\begin{figure}}
\newcommand{\eef}{\end{figure}}
\newcommand{\leg}[1]{\caption{\protect\rm{\protect\footnotesize{#1}}}}
\newcommand{\ew}[1]{\langle{#1}\rangle}
\newcommand{\be}[1]{\mid\!{#1}\!\mid}
\newcommand{\no}{\nonumber}
\newcommand{\etal}{{\em et~al }}
\newcommand{\geff}{g_{\mbox{\it{\scriptsize{eff}}}}}
\newcommand{\da}[1]{{#1}^\dagger}
\newcommand{\cf}{{\it cf.\/}\ }
\newcommand{\ie}{{\it i.e.\/}\ }

\title{Transport Through Nanostructures with Asymmetric Coupling to the Leads}

\author{G.~L.~Celardo}
\affiliation{Tulane University, Department of Physics, New Orleans, Louisiana 70118, USA}
\affiliation{Dipartimento di Matematica e Fisica, Universit\`a Cattolica, via Musei 41, 25121, Brescia, Italy}
\affiliation{I.N.F.N., Sezione di Pavia, Italy}
\author{A.~M.~Smith}
\affiliation{Tulane University, Department of Physics, New Orleans, Louisiana 70118, USA}
\author{S.~Sorathia}
\affiliation{Instituto de F\'{\i}sica, Universidad Aut\'{o}noma de
Puebla, Apartado Postal J-48, Puebla, Pue., 72570, M\'{e}xico}
\author{V.~G.~Zelevinsky}
\affiliation{NSCL and Department of Physics and Astronomy, Michigan State
University, East Lansing, Michigan 48824-1321, USA}
\author{R. A. Sen'kov}
\affiliation{Department of Physics, Central Michigan University, Mount Pleasant,
Michigan 48859, USA}
\author{L.~Kaplan}
\affiliation{Tulane University, Department of Physics, New Orleans, Louisiana 70118, USA}

\begin{abstract}

Using an approach to open quantum systems based on the effective
non-Hermitian Hamiltonian, we fully describe
transport properties for a paradigmatic model of a
coherent quantum transmitter: a finite sequence of square
potential barriers.
We consider the general case of asymmetric external barriers
and variable coupling strength to the environment.
We demonstrate that transport properties are very sensitive to
the degree of opening of the system and
determine the parameters for maximum transmission at any given degree
of asymmetry.
Analyzing the complex eigenvalues of the non-Hermitian Hamiltonian,
we show a double transition to a super-radiant regime where the
transport properties and the structure of resonances
undergo a strong change.
We extend our analysis to the presence of
disorder and to higher dimensions.
\end{abstract}

\date{\today}
\maketitle

\section{Introduction}

Open quantum systems, which exchange matter and energy with an environment, are
at the center of many research areas in condensed matter, atomic, molecular,
and nuclear physics. Major problems of current interest range from quantum
computing to transport in mesoscopic systems to basic theoretical issues,
including the measurement problem in quantum mechanics.

The nature and degree of opening affects the properties of a system in a highly nontrivial
manner. An example of this is the {\sl super-radiance} phenomenon in a finite
quantum system coupled to an environment characterized by a continuum of
states. Generically, at weak coupling, all internal states are similarly
affected by the opening and acquire small decay widths, resulting in narrow
transmission resonances. As the coupling increases and reaches a critical
value, the resonances overlap, and a sharp restructuring of the system occurs.
Beyond this critical value, a few resonances become short-lived states, leaving
all other (long-lived) states effectively decoupled from the environment. This
general phenomenon is referred to as the super-radiance
transition~\cite{SZNPA89,Zannals}, due to its analogy with Dicke
super-radiance~\cite{dicke54} in quantum optics.

In a recent work~\cite{Kaplan} generalizing the schematic tight-binding model
discussed in~\cite{Zannals,VZWNMP04}, it was shown that the phenomenon of
super-radiance actually occurs in the problem of transport through realistic
nanosystems. The specific situation analyzed in Ref.~\cite{Kaplan} is transport
through a one-dimensional ($1d$) sequence of potential barriers, see
Fig.~\ref{PWELL}. This paradigmatic model of solid state physics appears in
many important applications, including semiconductor superlattices and
one-dimensional quantum dot arrays. It has been widely discussed in the
literature~\cite{POT,TsuEsaki,fabry-Perot}; the transport properties have been
analyzed as the system--environment coupling was varied by adjusting the widths
of the external barriers. In Ref.~\cite{Kaplan} only {\sl symmetric} coupling
was considered, i.e. equal left and right external barriers. It was shown that
maximum transmission through the array is reached precisely at the
super-radiant transition. The transport properties of a $1d$ sequence of
potential barriers were analyzed with  the aid of the energy-independent {\sl
effective non-Hermitian Hamiltonian}. This approach produces excellent
agreement with  the exact (numerical) treatment of the problem for weak
tunneling between the wells.

In this paper we use the same framework to analyze the transport properties of a
$1d$ sequence of potential barriers with  {\it asymmetric} coupling to the
environment. We vary the external coupling strength, keeping the ratio between left and
right coupling constant. This allows us to determine the maximum transmission
as a function of the asymmetry and to extend the analysis to higher-dimensional
systems: quasi-$1d$, $2d$, and $3d$.

The paper is organized as follows. In Sec.~\ref{sec_ham}, we introduce the
energy-independent effective Hamiltonian that relates the sequence of barriers
to the open Anderson model, and show how the transmission properties can be
determined in this formulation. In Sec.~\ref{sec_maximum}, we find the strength
of coupling to the environment at which both  integrated transmission and
average transmission at the center of the energy band are maximized, and derive
the scaling of this critical coupling strength as a function of asymmetry and
degree of disorder. The structure of resonances is analytically computed as a
function of energy and of asymmetry of the coupling for the case of no
disorder. Analyzing the complex eigenvalues of the non-Hermitian Hamiltonian in
Sec.~\ref{sec_resonances}, a double transition to a super-radiant regime is
shown to occur. Contrary to the symmetric coupling case, the maximum
transmission in the asymmetric case is not reached at either transition, but
occurs instead at a critical value located between the two. The super-radiant
transitions have significant consequences for the transport properties of the
$1d$ Anderson model. The number of resonances decreases by one every time a
super-radiant transition is crossed. For a large number of sites in the $1d$
chain, a clear signature of the two super-radiant transitions is observed when
we analyze the resonance structure near the center of the energy band as a
function of the coupling strength  to the leads. In Sec.~\ref{sec_RMT}, we
compare our results with the random matrix theory of transport, showing that
random matrix theory is only partially applicable to the $1d$ Anderson model.
Finally, in Sec.~\ref{SecMD}, we extend our results to multi-dimensional
scenarios, showing that the validity of our findings is not limited to the $1d$
case.

\section{Effective Hamiltonian}
\label{sec_ham} The effective Hamiltonian approach to open quantum systems was
formulated in the book~\cite{MW} for nuclear reactions, and later generalized
and studied in detail, see for example~
\cite{SZNPA89,Zannals,VZWNMP04,rotter91,DittesR,Kaplan}. In Ref.~\cite{Kaplan}
two of the present authors demonstrated how to build an effective non-Hermitian
Hamiltonian that correctly describes transport through a sequence of potential
barriers. Here we will only review the main points of the approach and
establish definitions and notations.

\begin{figure}[h!]
\vspace{0cm}
\includegraphics[width=7cm,angle=-90]{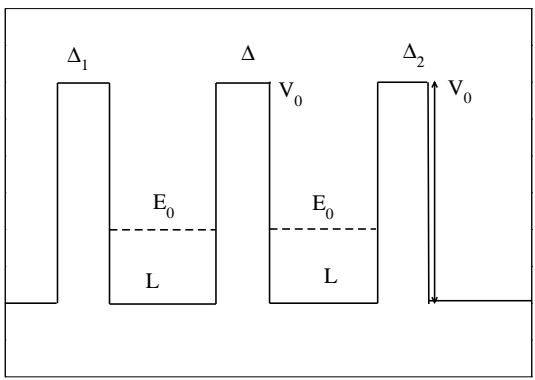}
\caption{Sequence of potential barriers of finite height and width.}
\label{PWELL}
\end{figure}

In the absence of disorder, we consider quantum transport through
a sequence of $N+1$ potential barriers
of height $V_0$ and inter-barrier separation $L$, as illustrated
in Fig.~\ref{PWELL}. All $N-1$ internal barriers have width $\Delta$,
while the two external barriers have widths $\Delta_{1,2}$.

We may compute transmission through the system in a standard way, by
matching the wave function and its derivative on either side of each barrier in
Fig.~\ref{PWELL}. Throughout the paper, we will use units with
$\hbar^2/2m_e=1$. Thus, when distances $\Delta$, $\Delta_{1,2}$, and $L$ are
measured in nm (the typical scale for semiconductor superlattices), all
energies are calculated in units of $38$ meV. In what follows we set $L=2$,
$\Delta=0.2$, $V_0=1000$, and $E_0=V_0/2=500$ (so that the energy shift
vanishes, see below).

In the limit of weak tunneling between the sites, a sequence of $N$ potential
wells behaves as an open $1d$ Anderson model. As shown in Ref.~\cite{Kaplan},
the $1d$ effective Hamiltonian for an energy band centered at $E_0$ can be
written in the site basis in a way similar to that used in
Refs.~\cite{Zannals,VZWNMP04},
\begin{equation}
{\cal H} =  \left( \begin{array}{ccccc}
E_0+\delta_1-\frac{i}{2} \gamma_1 & \Omega & 0 & ...& 0 \\
\Omega & E_0 & \Omega & ... & 0\\
0 & \Omega & E_0 & ... & 0 \\
...&...&...&...&...\\
0& 0 & 0& ... & E_0+\delta_2-\frac{i}{2} \gamma_2 \end{array} \right) \,.
\label{Heff}
\end{equation}
The edge states, $|1\rangle$, localized in the first well on the left, and
$|N\rangle$, localized in the last well on the right, acquire finite widths,
$\gamma_{1,2}$, and energy shifts, $\delta_{1,2}$, due to the coupling to the
environment. By inter-site tunneling, this coupling propagates through the
chain.

The effective Hamiltonian ${\cal H}$ correctly reproduces transmission
through a sequence of potential barriers if we define the tunneling coupling
$\Omega$ as
\begin{equation}
\label{Omega}
\Omega=   \frac{2 \alpha^2 E_0}{V_0 (1+\alpha L/2)} \exp{(-\alpha \Delta)} \,,
\end{equation}
where $\alpha=\sqrt{V_0-E_0}$. Similarly, the widths and energy shifts can be written as
\begin{equation}
\left. \begin{array}{lll}
         \gamma_{1,2} = \frac{8 \alpha^3 E_0 k}{V_0^2 (1+\alpha L/2)} \exp{(-2 \alpha \Delta_{1,2})},\\ \\
         \delta_{1,2} =  \frac{k^2-\alpha^2}{4 \alpha k}\, \gamma_{1,2}\,, \\
        \end{array} \right. \label{pars}
\end{equation}
where  $k=\sqrt{E_0}$. The shifts $\delta_{1,2}$ vanish for $E_0=V_0/2$;
otherwise the sign of $\delta$ is given by the sign of $E_0-V_0/2$. We will
study how the transport properties depend on the system--environment couplings
$\gamma_{1,2}$, which are varied by adjusting the external barrier widths
$\Delta_{1,2}$ while keeping all other parameters fixed.

With  the aid of the effective Hamiltonian, the transmission coefficient
$T^{ab}(E)$ from channel $b$ to channel $a$ can be determined,
\begin{equation}
T^{ab}(E)=|Z^{ab}(E)|^2 \,,
\label{T1}
\end{equation}
where
\begin{equation}
Z^{ab}(E)=\sum_{i,j=1}^N A_i^a \left( \frac{1}{E-{\cal H}}\right)_{i,j}(A_j^b)^\ast
\label{T2}
\end{equation}
is the transmission amplitude. The channels are labeled by the quantum numbers
that characterize the continuum states in the environment, not including the
energy. In the $1d$ case we have two channels, $a=1,2$, corresponding to the
left and right scattering states, respectively. The factors $ A_i^a$ represent
the transition amplitudes from state $|i\rangle$ to channel $a$. In our
arrangement, the only non-vanishing transition amplitudes are $ A_1^1 =
\sqrt{\gamma_1}$ and $A_N^2 = \sqrt{\gamma_2}$. The complex eigenvalues ${\cal
E}_{k}$ of ${\cal H}$ coincide with  the poles of $Z(E)$. The spectrum of the
complex eigenvalues of the effective Hamiltonian is of great importance for
understanding the transport properties of the system.

Using the effective Hamiltonian of Eq.~(\ref{Heff}), the transmission between
left and right leads can be computed as
\begin{equation}
T(E)= \left| \frac{(\sqrt{\gamma_1 \gamma_2}/ \Omega)}{ \prod_{k=1}^N (E-{\cal E}_k)/ \Omega} \right |^2 \,,
\label{THeff}
\end{equation}
where here and in the following $T(E) \equiv T^{12}(E)$. It was shown in Ref.
\cite{Kaplan} that the exact transmission obtained by matching of the wave
functions is in excellent agreement with  the effective Hamiltonian approach,
Eq.~(\ref{THeff}), for $\alpha \Delta \gg 1$.

\section{Transmission Through Nanostructures with  Asymmetric Coupling}
\label{sec_maximum}

In this Section we analyze the behavior of the transmission as we increase the
coupling to the external environment, keeping the ratio of the two external
couplings,  $\gamma_1=\gamma$ and $\gamma_2=\gamma/q$, fixed and equal to $q$,
the asymmetry parameter. We treat first the case of no disorder, where the
unperturbed states in all sites have the same energy $E_0$, and later extend to
the disordered case. We will also drop the energy offset $E_0$ from the
effective Hamiltonian, so that the center of the energy band is always at zero
energy, and for simplicity we neglect the energy shifts $\delta_{1,2}$ in the
following considerations (which in the potential of Fig.~\ref{PWELL}
corresponds to $E_0=V/2$).

\subsection{Ordered Case}
\label{orderedcase}

Before considering the problem of a general $N$-level system, we first treat
the one-well and two-well cases, which give us valuable insight into how
transport properties change as we increase the coupling strength to the environment.

We start with the transmission through a {\sl single quantum level}, namely
through a quasistationary state created by two potential barriers. In this case
we have only one resonance and the effective Hamiltonian reduces to ${\cal
H}=-i \gamma/2 -i \gamma/2q$. The resonance height is independent of the
coupling $\gamma$, and transmission is never perfect for asymmetric
barriers. Indeed, from Eq.~(\ref{THeff}) we see that maximum transmission is
attained at $E=0$ and is given by
\begin{equation}
T(E=0)=\frac{4q}{(q+1)^2} \,,
\label{TN1}
\end{equation}
so that $T(E=0)=1$ only for the equal-coupling case of $q=1$.

The situation is different when we consider transmission through {\sl two
quantum states}. This problem has been studied previously, see for instance
Refs.~\cite{VZ03,nonthermal} and references therein. In fact, many of the
results obtained for the two-level case are of more general validity. The
effective Hamiltonian for two originally degenerate levels can be written as
\begin{equation}
{\cal H}=  \left( \begin{array}{cc}
-\frac{i}{2} \gamma & \Omega \\
\Omega & -\frac{i}{2} \gamma/q \end{array} \right) \,.
\label{Heff2}
\end{equation}
From Eq.~(\ref{THeff}) we can then compute the transmission. In particular, the
transmission at $E=0$ is
\begin{equation}
T(E=0)= \left|\frac{(\gamma/ \Omega)/\sqrt{q} }{1+(\gamma/ \Omega)^2/4q}\right|^2 \,.
\label{T0}
\end{equation}
At the critical value,
\begin{equation}
\left(\frac{\gamma}{\Omega}\right)_{\rm cr}=2 \sqrt{q} \,,
\label{gcr3}
\end{equation}
transmission is perfect. This is in contrast with  the one-level situation,
where transmission is never perfect for $q \ne 1$.

By applying the residue method to Eq.~(\ref{THeff}), we can also compute the
normalized integrated transmission:
\begin{equation}
S=
\frac{1}{4 \Omega}\int  T(E) \, dE
=
\frac{\pi \gamma/ \Omega}{2(q+1)[1+(\gamma/ \Omega)^2/4q]} \,,
\label{STheory}
\end{equation}
where $4\Omega$ is the width of the energy band. The maximum integrated
transmission occurs at the same critical $\gamma$ value given by
Eq.~(\ref{gcr3}). The quantitative behavior of $S$ is important in
applications, for instance in the design of electron band-pass filters for
semiconductor superlattices~\cite{POT}.

\begin{figure}[h!]
\vspace{0cm}
\includegraphics[width=7cm,angle=-90]{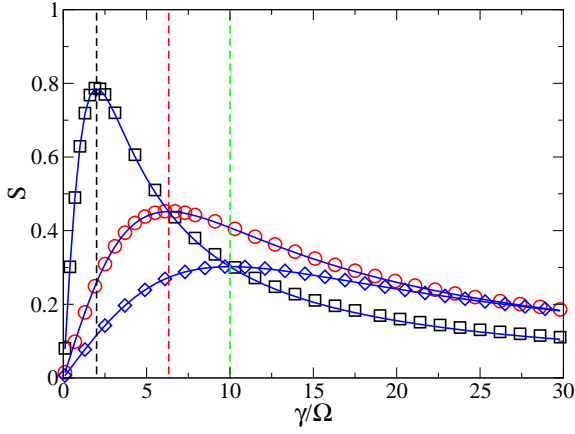}
\caption{ (Color online) Integrated transmission $S$ obtained in the effective
Hamiltonian approach for $N=100$ sites and different asymmetry parameters $q$:
squares refer to $q=1$, circles to $q=10$, and diamonds to $q=25$. The solid
curves represent the result (\ref{STheory}) obtained for $N=2$. The vertical
dashed lines indicate the critical value of the coupling strength in each case,
$\gamma/ \Omega = 2 \sqrt{q}$, where $S$ is predicted to have its maximum, see
Eq.~(\ref{gcr3}). } \label{AsymS}
\end{figure}

We now analyze transport properties for the general {\sl $N$-level system},
many of which parallel those of the special one- and two-level cases.
Numerically calculating transmission as a function of energy, we find that
perfect transmission is attained at all resonance peaks precisely at the
critical coupling given by Eq.~(\ref{gcr3}) independently of the value of $N$,
see for example the upper right panel of Fig.~\ref{Asym100T}. Moreover, for all
$N \ge 2$, the integrated transmission has a maximum at the same value of the
coupling. The simple theoretical expression (\ref{STheory}), obtained for the
integrated transmission in the two-level case, closely reproduces the
integrated transmission for any $N \ge 2$, see Fig.~\ref{AsymS}. The origin of
this $N$-independent behavior in the coherent transport regime has been
discussed previously in the context of symmetric coupling~\cite{Kaplan}.

Our results agree with the full analytical expression for the transmission
amplitude~(\ref{T2}) in the general $N$-level case~\cite{Zannals},
\begin{equation}
Z(E)=-\,\frac{2\sqrt{\gamma_{1}\gamma_{2}}\,P_{-}(E)}{1+i(\gamma_{1}+\gamma_{2})P_{+}(E)
+\gamma_{1}\gamma_{2}[P_{-}^{2}(E)-P_{+}^{2}(E)]},           \label{a}
\end{equation}
where
\begin{equation}
P_{\pm}(E)=\,\frac{1}{N+1}\sum_{n=1}^{N}(\pm 1)^{n}\,\frac{1}{E-E_{n}}\,
\sin^{2}\left(\frac{\pi n}{N+1}\right),                   \label{c}
\end{equation}
and the sums are over the unperturbed Bloch wave energies of the closed system,
\begin{equation}
E_{n}=2\Omega\cos\left(\frac{\pi n}{N+1}\right).        \label{b}
\end{equation}
This determines the exact transmission probability,
\begin{equation}
T(E)=\,\frac{4\gamma_{1}\gamma_{2}P_{-}^{2}}{\Bigl[1+\gamma_{1}\gamma_{2}
\Bigl(P_{-}^{2}-P_{+}^{2}\Bigr)\Bigr]^{2}+(\gamma_{1}+\gamma_{2})^{2}P_{+}^{2}}.
                                                                \label{d}
\end{equation}
Inside the energy band, at any pole $E=E_{n}$ corresponding to a Bloch
eigenstate, both sums $P_{\pm}$ diverge with $|P_+/P_-| \to 1$, and the
transmission takes the $N$-independent and $\gamma$-independent value given by
Eq.~(\ref{TN1}),
\begin{equation}
T(E=E_n)=T_1 \equiv \frac{4q}{(q+1)^2} \,,
\label{tpole}
\end{equation}
obtained above for $E=0$ in the special case $N=1$. Outside the band, for
$|E|>2\Omega$, the sum $P_{+}$ converges at large $N$ to the $N$-independent
value
\begin{equation}
P_{+}(\epsilon) \approx \frac{z_{\mp}}{2 \Omega}
\label{pplusoutside}
\end{equation}
(see Appendix), whereas the sum $P_{-}$ is exponentially small at large $N$,
\begin{equation}
P_{-}(\epsilon) \approx \mp \frac{\sqrt{\epsilon^2-1}}{\Omega z_{\pm}^{N+1}} \,,
\label{pminusoutside}
\end{equation}
where $\epsilon=E/2\Omega$ and $z_{\pm}=\epsilon \pm \sqrt{\epsilon^2-1}$. The
fast decay of $P_{-}$ results in exponentially weak transmission. In
Eqs.~(\ref{pplusoutside}) and (\ref{pminusoutside}), the upper and lower signs
should be chosen for $\epsilon > 1$ and $\epsilon < -1$, respectively. For
energies $E$ inside the band, the sums in Eq.~(\ref{c}) take a simple form, see
Appendix:
\begin{equation}
P_{\pm}(\epsilon) \approx \frac{1}{2 \Omega}
\begin{cases}  \frac{\sin (N\beta)}{\sin [(N+1)\beta]}\\
-\frac{\sin (\beta)}{\sin [(N+1)\beta]} \end{cases} \,, \label{exN}
\end{equation}
where $\beta=\cos^{-1}(\epsilon)$.

A convenient simplification does occur for energies near the middle of the
band, $|E| \ll \Omega$, where for large $N$ the sums $P_{\pm}$ are dominated by
terms associated with $n \approx N/2$, while distant contributions from the
left and right sides of the Bloch spectrum cancel. Near the center of the band,
$E_n$ may be replaced by a picket fence spectrum with spacing $D=2 \pi
\Omega/N$, while the last factor in Eq.~(\ref{c}) reduces to unity. Defining
$r$ by $E=E_{n_0}+r D$, we then have the $N-$independent result
\begin{equation}
P_{\pm}(E) \approx \frac{1}{2 \Omega}
\begin{cases}  \cot (\pi r) \\   (\pm 1)^{n_0} \csc (\pi r) \end{cases} \,,
\label{r}
\end{equation}
and therefore
\begin{equation}
T(E) \approx \frac{16 \gamma_1 \gamma_2 }
{(4\Omega+\gamma_1\gamma_2/\Omega)^2\sin^2(\pi r)+4 (\gamma_1+\gamma_2)^2\cos^2(\pi r)}
\,.
\label{tbandcenter}
\end{equation}
At energies in the Bloch spectrum ($r=0$), Eq.~(\ref{tbandcenter}) reduces, as
it must, to the exact expression (\ref{tpole}), while midway between any two
neighboring poles ($r=1/2$) the transmission becomes
\begin{equation}
T(E=E_n+D/2) \approx T_2 \equiv \left|\frac{(\gamma/ \Omega)/\sqrt{q}
}{1+(\gamma/ \Omega)^2/4q}\right|^2 \,, \label{t2def}
\end{equation}
which agrees with the transmission at $E=0$ obtained above in the special case
of $N=2$ wells, Eq.~(\ref{T0}). At these energy midpoints, perfect transmission
occurs at the critical value of the coupling given by Eq.~(\ref{gcr3}), just as
it does in the special case $N=2$. Averaging Eq.~(\ref{tbandcenter}) over an
energy window containing multiple resonances, $\Omega/N \ll \Delta E \ll
\Omega$, we obtain the energy-averaged transmission near the middle of the
band,
\begin{equation}
\overline{T}=\int_0^1 T(E) \,dr \approx \frac{8 \gamma/\Omega }{(q+1)(4+(\gamma/\Omega)^2/q)} \,,
\label{tavgmiddle}
\end{equation}
that reaches its maximum, $T_{\rm avg}^{{\rm max}}=2\sqrt{q}/(1+q)$, again at
the critical value of the coupling $\gamma/\Omega$ given by Eq.~(\ref{gcr3}).

The critical value associated with the maximum transmission in Fig.~\ref{AsymS}
is also consistent with the results of Ref.~\cite{PB}. There, the authors found
that given a sequence of potential barriers of width  $\Delta$, adding two
external barriers satisfying $\Delta_L+\Delta_R=\Delta$ produces an increase in
transmission while leaving the resonance energies unchanged. From
Eqs.~(\ref{Omega}) and (\ref{pars}), we can see that the maximum transmission
condition (\ref{gcr3}) obtained from the tight-binding model coincides with the
condition $\Delta_L+\Delta_R=\Delta$ for the case of $E_0=V_0/2$, in which case
the energy shift is zero.

\subsection{Disordered Case}

In this subsection we analyze the effect of disordered on-site energies. The
survival of the super-radiant restructuring in the disordered chain was
established in Ref. \cite{VZWNMP04}. In Ref.~\cite{Kaplan} we showed that the
effective Hamiltonian, Eq.~(\ref{Heff}), correctly describes the sequence of
potential barriers in the presence of disorder (e.g., variable well width),
when the disorder is sufficiently weak. We consider  random variations of the
diagonal energies, $E_0 + \delta E_0$, where $\delta E_0$ is uniformly
distributed in the interval $[-W/2 ,+W/2]$, and $W$ is the disorder parameter.
In Ref.~\cite{Kaplan} it was shown that for $\alpha \Delta \gg 1$, a random
variation of $\delta E_0$ in the interval $[-W/2,+W/2]$ corresponds to a random
variation of the well width  $\delta L$ in $[-WL/4E_0, +WL/4E_0]$.

The effective non-Hermitian Hamiltonian with  diagonal disorder is equivalent
to a $1d$  open Anderson tight-binding model~\cite{Anderson,Lee}. The
eigenstates of the Anderson model are exponentially localized on the system
sites, with the tails given by $\exp(-x/L_{\rm loc})$. Here the localization
length  $L_{\rm loc}$ depends on the disorder strength~\cite{Felix}, and $x$ is
distance in the direction of transmission, measured in units of the well size.
For $L_{\rm loc}\ll N$, the transmission decays exponentially with  $N$; this
is the localized regime.

\begin{figure}[h!]
\vspace{0cm}
\includegraphics[width=7cm,angle=-90]{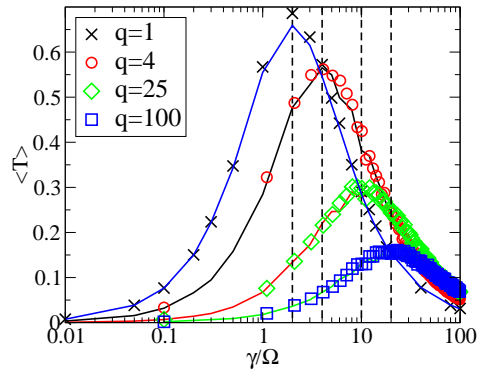}
\caption{ (Color online) Average transmission near $E=0$ as a function of
coupling strength  $\gamma/ \Omega$ for a disordered $1d$ chain of $N=100$
wells, for different values of the asymmetry parameter $q$. The disorder
strength  is $W/\Omega=0.5$. Each value of transmission corresponds to the
average over an energy window $-0.1<E<0.1$, and ensemble average over $100$
realizations of the disorder. The curves are obtained from the effective
Hamiltonian ${\cal H}$ (open Anderson model), whereas symbols are computed for
the sequence of potential barriers. Vertical dashed lines indicate the critical
values of the coupling (\ref{gcr3}) for the ordered case. } \label{Asym1D}
\end{figure}

Let us first consider the case when the mean level spacing $D$ (at the center
of the energy band) is not strongly modified by the disorder. In the ordered
case, $D \approx 2 \pi \Omega/N$ for $E=0$, while for strong disorder the band
width is $W$ and we have $D \approx W/N$. Thus, we expect that for $W < 2 \pi
\Omega$ the mean level spacing is not strongly influenced by the disorder. This
regime is shown in Fig.~\ref{Asym1D}, where we plot the average transmission as
a function of the coupling strength  $\gamma/ \Omega$ for $N=100$ and
$W/\Omega=0.5$. As indicated by the vertical dashed lines, the maximum
transmission is reached at the same critical values of $\gamma/ \Omega$
obtained previously in the absence of disorder, Eq.~(\ref{gcr3}). At this
critical value, each transmission curve intersects the transmission curve for
the symmetric case $q=1$. Indeed, at the critical coupling, the tunneling
probabilities from the left and right are equal, implying that for this value
the asymmetric system behaves as a system with  symmetric coupling, see
Sec.~\ref{sec_RMT}.

\begin{figure}[h!]
\vspace{0cm}
\includegraphics[width=7cm,angle=-90]{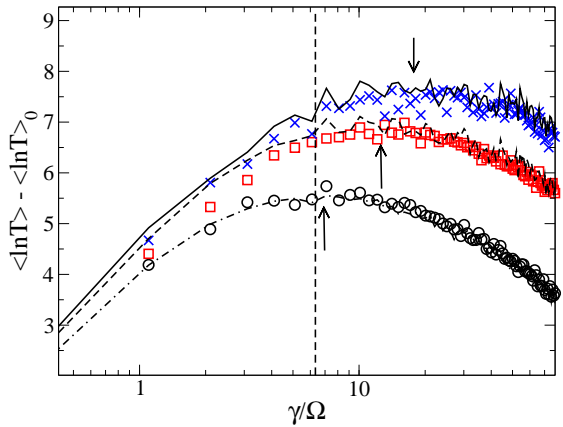}
\caption{ (Color online) Average of the logarithm of transmission at $E=0$ as a
function of the coupling $\gamma/ \Omega$ for the $1d$ case, with $N=100$
sites, and asymmetry parameter $q=10$. In our units, the tunneling coupling is
$\Omega=1$, and the three data sets correspond to different disorder strengths
$W$. $\langle \ln T \rangle_0$ represents the value of the transmission at
$\gamma/ \Omega = 0.01$, which is the lowest value of  $\gamma/ \Omega$
computed. For the sake of clarity, we plot $\langle \ln T \rangle-\langle \ln T
\rangle_0$ so that all data start from the same point. In each case, symbols
refer to the results obtained from the effective Hamiltonian ${\cal H}$:
circles for $W=4$, squares for $W=10$, and crosses for $W=15$, while the curves
refer to a sequence of potential barriers: dot-dashed for $W=4$, dashed for
$W=10$, and solid for $W=15$. Averaging is performed over $1000$ realizations
of the disorder. The dashed vertical line indicates the critical value of the
coupling, Eq.~(\ref{gcr3}), at which transmission is maximized for low
disorder. The arrows indicate $\gamma/\Omega= 1.2 ND/\Omega $, where $D$ is the
mean level spacing at the center of the energy band and the factor $1.2$ is
obtained from fitting. } \label{surenW}
\end{figure}

The situation is different for strong disorder, $W> 2 \pi \Omega$. The critical
value for maximized transmission now depends on the disorder strength. For
strong disorder we enter the localized transport regime studied in
Ref.~\cite{Beenakker}, where transmission is log-normally distributed, so
that it is more convenient to consider the average of $\ln T$ rather than the
average of $T$. Numerically we find that the value at which transmission is
maximized is proportional to the mean level spacing $D$ at the center of the
energy band: $\gamma \propto ND $, see the arrows that indicate $\gamma =1.2 ND
$ for each curve in Fig.~\ref{surenW}, where $q=10$. Since  $D \approx W/N$
when disorder is strong, the  critical value of the coupling is in this case
proportional to the disorder strength.

Additionally, we note the close agreement between symbols and solid curves in
Figs.~\ref{Asym1D} and \ref{surenW}, demonstrating good correspondence between
the exact barrier problem, Fig.~\ref{PWELL}, and the energy-independent
effective Hamiltonian of Eq.~(\ref{Heff}). This correspondence persists for
weak and strong coupling to the environment, in highly symmetric and highly
asymmetric situations, and also for both weak and strong disorder.

\section{Double Super-Radiant Transition and Structure of
Resonances}

\label{sec_resonances}

\begin{figure}[h!]
\vspace{0cm}
\includegraphics[width=7cm,angle=-90]{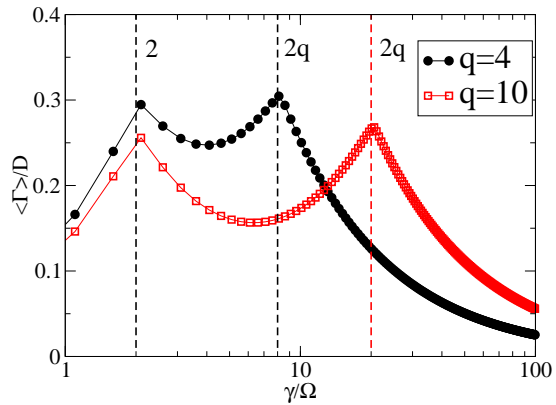}
\caption{ (Color online) Average width  (computed over the smallest $N-2$
widths) in units of the mean level spacing at the center of the energy band.
The average width is calculated as a function of $\gamma/\Omega$ for $N=100$
and two values of the asymmetry parameter $q$. Dashed vertical lines indicate
the locations of the two super-radiant transitions, as given by
Eq.~(\ref{2SR}). } \label{AsymGamma}
\end{figure}

In Ref.~\cite{Kaplan} we showed that maximum transmission is reached at the
super-radiant transition in the symmetric coupling case. The transition is
signaled by a segregation of resonance decay widths, i.e., of the imaginary
parts of the eigenvalues of the effective Hamiltonian, into two groups,
super-radiant (short-lived) and trapped
(long-lived)~\cite{SZNPA89,Zannals,Kaplan,izr94}. The number of super-radiant
states is equal to the number of channels coupling the system to the
environment [two for Fig.~\ref{PWELL} and for the effective Hamiltonian of
Eq.~(\ref{Heff})]. In order to identify the super-radiance transition we
compute the average of the non-super-radiant widths, i.e. of the smallest $N-2$
decay widths of the effective Hamiltonian. In Fig.~\ref{AsymGamma}, we plot
this average value as a function of the coupling $\gamma/ \Omega$ for two
asymmetry values, $q=4$ and $q=10$. In each case the average width  shows two
maxima as $\gamma / \Omega$ is varied. At each maximum, one of the widths
segregates from the others. Thus, in the presence of asymmetry, we have two
super-radiant transitions associated with two critical values of the coupling.
According to the resonance overlap condition, $\gamma_{1,2}/
\Omega=2$~\cite{Kaplan}, the two transitions are predicted to occur at the
values
\begin{equation}
(\gamma/\Omega)_{\rm SR1}=2, \hspace {0.5cm}  \hspace {0.2cm}(\gamma/\Omega)_{\rm SR2}=2q,
\label{2SR}
\end{equation}
indicated by the vertical dashed lines in Fig.~\ref{AsymGamma}, in very good
agreement with the numerical results for different values of the asymmetry.

Comparing Eqs.~(\ref{gcr3}) and (\ref{2SR}), we observe that, in the presence
of asymmetry, the point of maximum transmission is always in between the two
super-radiant transitions. The transmission maximum occurs approximately at the
minimum of the average decay width, while the super-radiant transitions occur
at the two maxima of the average decay width. This can be compared to the case
of symmetric barriers, where the maximal transmission precisely coincides with
the (single) super-radiant transition~\cite{Kaplan}, in agreement with Eq.~(\ref{2SR})
for $q=1$.

Since the super-radiant states are very broad, the number of observed
resonances changes after each transition. As we increase the coupling, the
number of resonances changes from $N$ at weak coupling to $N-1$ after the first
transition, and finally to $N-2$ for strong coupling. This change in the
resonance structure can have important consequences for experimental
current-voltage curves, for example in semiconductor
superlattices~\cite{TsuEsaki}. For symmetric coupling, the number of resonances
changes directly from $N$ to $N-2$ at the (single) super-radiant
transition~\cite{Kaplan}.

\begin{figure}[h!]
\vspace{0cm}
\includegraphics[width=7cm,angle=-90]{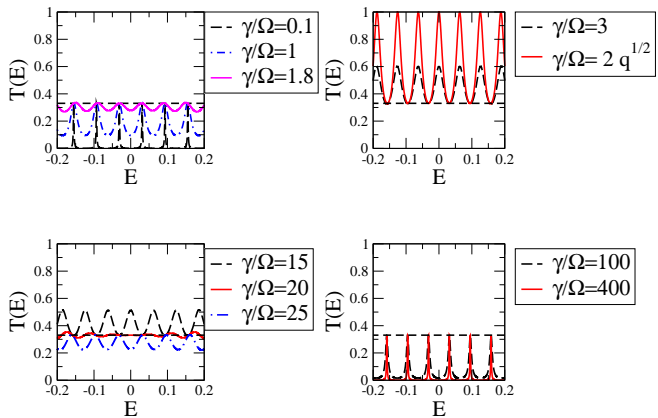}
\caption{ (Color online) Transmission for $N=100$ and asymmetry parameter
$q=10$ calculated as a function of energy in a small interval around the center
of the energy band. Ten different values of the coupling $\gamma/ \Omega$ are
shown for the case of no disorder, $W=0$. The horizontal dashed line,
Eq.~(\ref{tpole}), is the analytically obtained transmission at the location of
the poles, which is also the maximum transmission for $N=1$, Eq.~(\ref{TN1}). }
\label{Asym100T}
\end{figure}

An important consequence of the two super-radiant transitions can be seen in
the structure of resonances near the center of the energy band. In
Fig.~\ref{Asym100T} we consider a $1d$ chain with  $N=100$, $q=10$, and no
disorder ($W=0$), and calculate transmission as a function of energy for
different values of the coupling $\gamma/\Omega$. In each panel in
Fig.~\ref{Asym100T}, the dashed horizontal line indicates the transmission at
energy values belonging to the Bloch spectrum, Eq.~(\ref{tpole}), which is also
the maximal transmission attainable for transport through one level only, as
given by Eq.~(\ref{TN1}).

In the upper left panel of Fig.~\ref{Asym100T}, the resonance structure is
shown for several cases of weak coupling, $\gamma/ \Omega<2$, i.e. below the
first super-radiant transition. The maximum resonance height is here
independent of $\gamma/\Omega$, and equal to the single-level resonance height.
Only the widths grow as $\gamma/ \Omega$ is increased in this regime. In the
upper right panel, we consider $2< \gamma/ \Omega <2 q$, the intermediate
regime between the two super-radiant transitions. Perfect transmission is
reached for all resonances precisely at the geometric mean of the two
transitions, at  $\gamma/ \Omega=2 \sqrt{q}$. In this regime, it is the {\it
minimum} transmission that is independent of $\gamma/\Omega$ and equal to the
single-resonance transmission height,  Eq.~(\ref{TN1}). In the lower left panel
the region around the second super-radiant transition is shown. As $\gamma/
\Omega$ crosses the value $\gamma/ \Omega =2q$, the minimum transmission drops
below the one-level transmission value. Finally, in the lower right panel, we
show the regime of large coupling, $\gamma/ \Omega>2q$. Here it is again the
{\it maximum} transmission that is independent of $\gamma/\Omega$ and given by
Eq.~(\ref{TN1}), just as in the weak coupling regime, but the resonance widths
now {\it shrink} with  increasing $\gamma/ \Omega$. The behavior shown in
Fig.~\ref{Asym100T} indicates that, at least for large $N$, a qualitative
change in the resonance structure near the center of the energy band occurs at
each of the two super-radiant transitions.

\begin{figure}[h!]
\vspace{0cm}
\includegraphics[width=7cm,angle=-90]{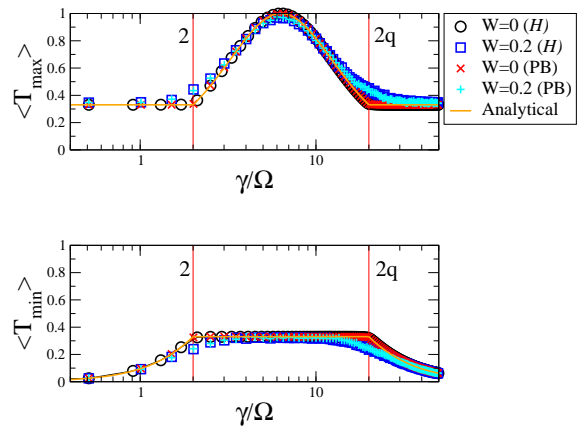}
\caption{ (Color online) The case of $N=100$ sites with  $q=10$ is considered.
Upper panel: the resonance height (transmission maximum) averaged over the
center of the band ($-0.1 <E<0.1$), as a function of $\gamma/ \Omega$. Lower
panel: the transmission minimum averaged over the same energy window. The two
vertical lines indicate the positions of the two super-radiant transitions. The
solid curve shows the predictions of Eqs.~(\ref{Tmax2}) and (\ref{Tmin2}). A
clear change in the transport properties is seen at each of the two
transitions. The effect persists in the presence of disorder ($W/\Omega=0.2$).
In each panel, the results obtained from the effective Hamiltonian (${\cal H}$)
are compared with those obtained from the sequence of potential barriers (PB).
} \label{Tmax}
\end{figure}

To better study this change in the resonance structure, we now compute $\langle
T_{\rm max} \rangle$, the average of the transmission maxima, and $\langle
T_{\rm min} \rangle$, the average of the transmission minima, as functions of
the coupling strength. In each case, the averaging is performed over a window
near the center of the energy band, $-0.1 < E <0.1$. The upper panel of
Fig.~\ref{Tmax} shows that $\langle T_{\rm max} \rangle$ is constant to the
left of the first super-radiant transition and to the right of the second
transition (shown as vertical lines). The constant value coincides with that
for the one-level maximal transmission, Eq.~(\ref{TN1}), as shown by the
horizontal line. Between the two transitions, $\langle T_{\rm max}\rangle$
reaches a maximum value at $\gamma/ \Omega= 2 \sqrt{q}$. In the lower panel of
Fig.~\ref{Tmax}, we show the behavior of $\langle T_{\rm min} \rangle$ as a
function of $\gamma/ \Omega$. This average rises to reach the one-level
transmission value at the first transition, stays constant up to the second
transition, and then decreases.

For the ordered case we then find numerically the simple results:
\begin{equation}
\langle T_{\rm max} \rangle = \left\{  \begin{array}{cc}
T_2 & {\rm for} \hspace{0.2cm} 2 \le \gamma/\Omega \le 2q \\
T_1& {\rm otherwise}
\end{array}  \right. \,,
\label{Tmax2}
\end{equation}
and
\begin{equation}
\langle T_{\rm min} \rangle = \left\{  \begin{array}{cc}
T_1 & {\rm for} \hspace{0.2cm} 2 \le \gamma/\Omega \le 2q \\
T_2 & {\rm otherwise} \end{array}
\label{Tmin2}  \right. \,,
\end{equation}
which agree well with the analysis in Sec.~\ref{orderedcase}. Here $T_1$ is the
transmission at resonant energies in the Bloch spectrum, Eq.~(\ref{tpole}),
that coincides with  the one-level maximal transmission value, Eq.~(\ref{TN1}).
Similarly, $T_2$ is the transmission value at midway points between these
energies, Eq.~(\ref{t2def}), that coincides with $T(E=0)$ for the two-level
case, Eq.~(\ref{T0}). These formulas, represented by the solid lines in
Fig.~\ref{Tmax}, are in good agreement with  the numerical calculation.

It is also interesting to point out that from the analytical expressions,
Eqs.~(\ref{t2def}) and (\ref{tpole}), we have that $T_2 \ge T_1$  for $2 \le
\gamma/\Omega \le 2q $, so that we regain the thresholds for the two
super-radiant transitions.

In the same Figure, we also illustrate the effect of adding disorder. For weak
disorder, the behavior is similar to the ordered case, while for strong
disorder the effect of the two super-radiant transitions is smoothed out. In
Fig.~\ref{Tmax} the results obtained from the effective Hamiltonian (circles
and squares) are compared with  the results from the sequence of potential
barriers (crosses and pluses). Again, the agreement is excellent, both  with
and without disorder.


\section{Comparison with  Random Matrix Theory}
\label{sec_RMT}

It is interesting to compare our results on the $1d$ Anderson model with
standard results obtained in the framework of the random matrix theory of
transport. In Ref.~\cite{Weiden}, the dependence of the conductance on the
degree of opening was analyzed for a quasi-$1d$ system with symmetric coupling
to the leads, while the case of asymmetric coupling was later addressed in
Ref.~\cite{Beenakker}, with different tunneling probabilities for the right and
left ends. The tunneling probability used in Ref.~\cite{Beenakker} corresponds
to the ``elastic scattering" probability $\tau$ defined in Ref.~\cite{puebla},
where the case of symmetric coupling and varying degree of internal disorder
was considered. This probability for channel $a$ is defined as  $\tau^a= 1-
|\langle S^{aa}\rangle|^2$, where $S$ is the scattering matrix,
$S^{ab}=\delta^{ab}-iZ^{ab}$, and $Z^{ab}$ is given by Eq.~(\ref{T2}). Again,
we take $a=1,2$ to be the channels corresponding to the left and right
scattering states, respectively. From Ref.~\cite{puebla} we know that in the
random matrix framework and at the center of the energy spectrum
\begin{equation}
\tau_{1,2}= \frac{4 \kappa_{1,2}}{(1 + \kappa_{1,2})^2} \,,
\label{tau}
\end{equation}
where $\kappa_{1,2}$ is the effective coupling to channel $1$ and $2$,
respectively. In the special case when the internal system is described by a
GOE random matrix,
\begin{equation}
\kappa_{1,2} =\frac{\pi \gamma_{1,2}}{2ND} \,,
\label{kappa}
\end{equation}
where $D$ is the mean level spacing at the center of the spectrum and $N$ is
the dimension of the internal system. As before, in the presence of asymmetry,
$\gamma_1=\gamma$ and $\gamma_2=\gamma/q$.

For asymmetric coupling, maximum transmission is achieved when $\tau_1 =
\tau_2$~\cite{Beenakker}. At this special point, the system with asymmetric
coupling behaves as a system with symmetric coupling. This happens in the
trivial case $\kappa_1 = \kappa_2 $ (where the coupling is symmetric to begin
with) or for $\kappa_1 =1/\kappa_2$. Since in the $1d$ case we have $ND \approx
2 \pi \Omega$ at the center of the energy band and for moderate disorder, we
find from Eqs.~(\ref{tau}) and (\ref{kappa}) that the maximum transmission
should be achieved when $\gamma/ \Omega= 4 \sqrt{q}$, which is off by a factor
of two from the value given by Eq.~(\ref{gcr3}).

In order to understand the origin of this difference, we rederive below
Eqs.~(\ref{tau}) and (\ref{kappa}) in a slightly different way, following the
approach of Refs.~\cite{puebla}. The scattering matrix  averaged over the
ensemble of random realizations is given by
\begin{equation}
\langle S\rangle=\frac{1-i\langle K\rangle}{1+i\langle K\rangle}
                                       \label{14}
\end{equation}
to leading order in $1/N$~\cite{SZNPA89,puebla}. Here $S^{ab}$ and $K^{ab}$ are
matrices in the channel space, with the explicit expression for $K^{ab}$ given
below in Eq.~(\ref{15}). Eq. (\ref{14}) is valid, under the assumption that the
internal system can be described by random matrix theory, when the elements of
$K$-matrices in the numerator and the denominator of Eq. (\ref{14}) are
effectively uncorrelated since their correlations lead only to corrections no
larger than $\sim 1/N$. In the $1d$ Anderson model, the above assumption breaks
down both in the limit of very weak disorder, where the internal system
approaches integrability, and also for extremely strong disorder, where we
enter the localized regime.

The $K$-matrix in channel space can be written as
\begin{equation}
K^{ab}(E)=\frac{1}{2}\,\sum_{n}\frac{
B^{a}_{n} B^{b\ast}_{n}}{E-E_{n}},
\label{15}
\end{equation}
where real energies $E_{n}$ are the eigenvalues of the closed system. In our
case of two channels, $B^{1}_{n}= \sqrt{\gamma_1} \langle 1|\psi_{n} \rangle$
and $B^{2}_{n}= \sqrt{\gamma_2} \langle N|\psi_{n} \rangle$ are the transition
amplitudes for the eigenstate $|\psi_{n}\rangle$ to the left and right leads,
respectively. The sum in Eq.~(\ref{15}) contains the eigenvalues of the
resolvent $1/(E-H)$ of the closed system. For energy $E$ inside the spectrum of
$H$ we should understand it as a limiting value, $E\rightarrow E+i0$. Using the
identity
\begin{equation}
\frac{1}{E-E_{n}+i0}={\rm P.v.}\,
\frac{1}{E-E_{n}}-i\pi\delta(E-E_{n}) \,,            \label{16}
\end{equation}
and replacing the summation with the integral, we can compute the  average
$K$-matrix. The principal value part is a smooth function of energy that
vanishes in the middle of the spectrum; as a result, in this vicinity
\begin{equation}
  \langle K^{11}\rangle=-\frac{i\pi}{2} \langle |B^{1}_{E_{n}=0}|^2 \rangle
  \rho(E=0)= -i \kappa_1 \,,
\label{17}
\end{equation}
where $\rho(E=0)=1/D$ is the density of states at the center of the
spectrum. Thus we have
\begin{equation}
\kappa_{1,2}= \frac{\pi |B^{1,2}_{E_{n}=0}|^2}{2D} \,,
\label{kappa2}
\end{equation}
which we can use in Eq.~(\ref{tau}). In particular, if the eigenstate
components are assumed to obey random matrix statistics, we have
$|B^{1,2}_{E_{n}=0}|^2=\gamma_{1,2} /N$, and we recover Eq.~(\ref{kappa}).

\begin{figure}[h!]
\vspace{0cm}
\includegraphics[width=7cm,angle=-90]{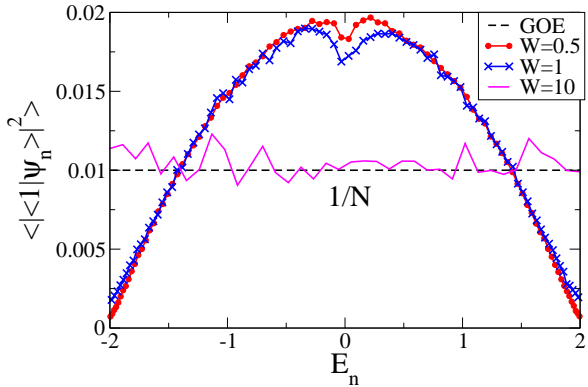}
\caption{ (Color online) Overlap probability of an eigenstate at energy $E_n$
with the edge $|1 \rangle$ of the chain in the $1d$ Anderson model for $N=100$.
The result is averaged over $10^4$ random realizations and plotted versus
energy for different disorder values $W$ (in units $\Omega=1$). The theoretical
value for a GOE random matrix is indicated by a dashed horizontal line. }
\label{AGOE}
\end{figure}

We analyze the statistics of the $1d$ Anderson model eigenstates in
Fig.~\ref{AGOE}, where  we plot  the ensemble-averaged value of $|\langle
1|\psi_{n}\rangle |^2$ (the probability overlap of the eigenstate
$|\psi_n\rangle$ with a state localized at the left edge of the chain, that for
a weakly open system would determine the width distribution for intrinsic
states) as a function of $E_{n}$ for different strengths of disorder.
Evidently, the components of the eigenstates do not follow random matrix theory
for moderate disorder. Near the center of the energy band we have $\langle
|\langle 1|\psi_{n}\rangle |^2 \rangle \approx 2/N$ for $W/\Omega=0.5$ and
$W/\Omega=1$. Then from Eq.~(\ref{kappa2}) we obtain
\begin{equation}
\kappa_{1,2} = \frac{\gamma_{1,2}}{2 \Omega } \,,
\label{kappa3}
\end{equation}
and $\gamma_{\rm cr}/ \Omega \approx 2 \sqrt{q}$, in agreement with our
findings in the previous sections. The values of $\gamma / \Omega$ at which we
have perfect tunneling probability,  $\gamma / \Omega= 2$, where $\tau_1=1$,
and $\gamma / \Omega= 2q$, where $\tau_2=1$, coincide precisely with the values
of $\gamma$ at which the two super-radiant transitions occur, see the
discussion in Sec.~\ref{sec_resonances}.  The fact that perfect tunneling
probability $\tau$ is reached at the super-radiant transition has been pointed
out in Refs.~\cite{puebla}.

From Fig.~\ref{AGOE} we see that for large values of $W$ we have $\langle
|\langle 1|\psi_{n}\rangle |^2 \rangle \approx 1/N$, but in this regime the
eigenstates are strongly localized and we no longer expect Eq.~(\ref{14}) to be
valid. The dip in the value of $\langle |\langle 1 |\psi_{n} \rangle|^2\rangle$
at the center of the energy band shown in Fig.~\ref{AGOE} is consistent with
the analysis of Ref.~\cite{Felix}, where it was pointed out that the
localization length is shorter at the center of the energy band than for
surrounding energies.

\begin{figure}[h!]
\vspace{0cm}
\includegraphics[width=7cm,angle=-90]{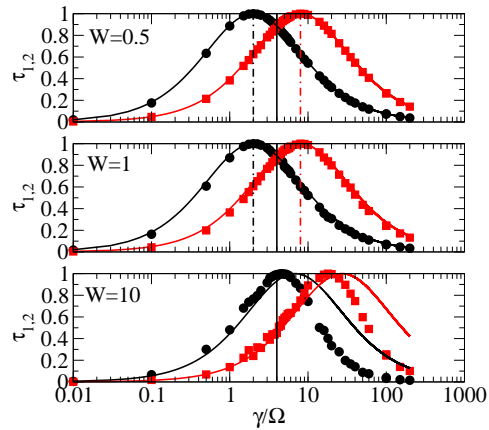}
\caption{ (Color online) Left and right tunneling probabilities $\tau_{1,2}$ are plotted
vs $\gamma/ \Omega$ for three different disorder strengths $W$, in units
$\Omega=1$, for the system size $N=100$ and asymmetry parameter $q=4$. The
numerical results (symbols) are obtained by averaging over $10^3$ disorder
realizations and over an energy interval $-0.1<E<0.1$ around the center of the
spectrum. These are compared with the theoretical results (solid curves)
obtained from Eqs.~(\ref{kappa3}) and (\ref{tau}). The critical value of
$\gamma/ \Omega$, Eq.~(\ref{gcr3}), is indicated by a vertical solid line in
each panel. In the upper and middle panels, the values at which perfect
transmission is reached, $\gamma/ \Omega=2$ and $\gamma/ \Omega=2q$, are shown
as vertical dot-dashed lines. } \label{1dtau}
\end{figure}

In Fig.~\ref{1dtau} we compare our numerical results for the tunneling probability
$\tau$ with Eqs.~(\ref{tau}) and (\ref{kappa3}), showing reasonable agreement for
moderate disorder $W/\Omega =0.5$, $1$. For $W/\Omega=10$, clear deviations
from the analytical expressions are visible, due to the fact that the
assumptions of random matrix theory break down at very strong disorder.

These results show that a blind application of random matrix results would lead
to incorrect conclusions for the $1d$ Anderson model. Our empirical expression
for the tunneling probability works well for moderate disorder. In this
regime, we regain the critical value of the coupling strength (\ref{gcr3}) for
which maximum transmission is achieved.


\section{Multi-Dimensional Case}
\label{SecMD}

In this Section, we discuss the higher-dimensional cases. Only selected results
will be shown, sufficient to demonstrate that the general behavior of the
maximum transmission found in $1d$ systems can be extended to higher
dimensions.

\begin{figure}[h!]
\vspace{0cm}
\includegraphics[width=7cm,angle=-90]{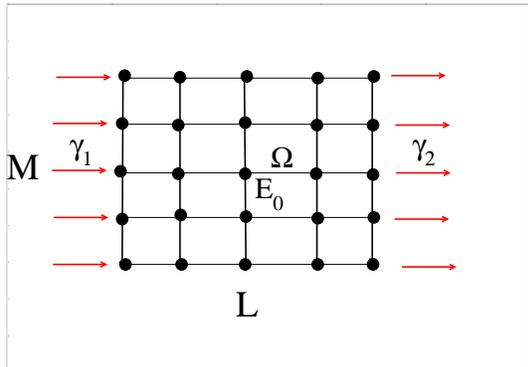}
\caption{ Two-dimensional open model used in this paper, with $N=M \times L$
sites coupled to $M$ incoming left channels and $M$ outgoing right channels.
Specifically, in this Figure we show $M \times L = 5 \times 5$ sites coupled to
$2M=10$ channels. At each site  there is a bound state with energy $E_0$
coupled to its nearest-neighbor sites through a tunneling amplitude $\Omega$.}
\label{Model}
\end{figure}

The open model in dimensions greater than one consists of an array of sites
with associated energy levels. Neighboring sites are coupled to each other by
the tunneling coupling $\Omega$, as in the $1d$ case. In higher dimensions, we
have many ways to couple the system to external leads. In the $2d$ case, where
the number of sites in a rectangular array is $N= M \times L$, we couple each
of the $M$ sites on the left side to a separate left lead and each of $M$ sites
on the right side to a separate right lead, see Fig.~\ref{Model}. The coupling
amplitude for each of the $M$ left leads is $\gamma_1=\gamma$, while the
coupling amplitude to each of the $M$ right leads is $\gamma_2=\gamma/q$.
Similarly, in the $3d$ case, with $N=M \times M \times L$ sites, each site on
an $M \times M$ face is coupled to a lead. In this geometry, each lead
represents a channel. Such an open model can describe a variety of physical
systems, such as an array of quantum dots, or a particle trapped in a lattice
potential.

The diagonal part of the effective Hamiltonian for this system can be written as
 ${\cal H}_{ii}=E_0 + \delta E_0$ for sites $i$ that are not coupled to leads, and
${\cal H}_{ii}=E_0 + \delta E_0+\delta_{1,2} -(i/2) \gamma_{1,2}$ for sites
coupled to the left or right leads, respectively. As in the $1d$ case, we set
the center $E_{0}$ of the energy band to zero; $\delta E_0$ is a random
variable uniformly distributed in $[-W/2 ,+W/2]$, and $W$ is a disorder
parameter. As before, $\delta_{1,2}$ and $\gamma_{1,2}$ represent the energy
shift and decay probability (inverse lifetime), respectively, induced by the
coupling to the left and right leads. In the following we again neglect the
energy shift, so that $\delta_{1,2}=0$. Finally, for the off-diagonal matrix
elements, $i\neq j$, we have ${\cal H}_{ij}={\cal H}_{ji}=\Omega$ if the sites
$i$ and $j$ are neighboring and ${\cal H}_{ij}=0$ otherwise.

Depending on the degree of disorder, different transport regimes are possible
in the multidimensional case. The ballistic regime is defined by the condition
$L<l$, where $L$ is the system length and $l$ is the mean free path. The
diffusive regime is determined by the condition $l<L<L_{\rm loc}$, where as
before $L_{\rm loc}$ is the localization length. Finally, for $L_{\rm loc}<L$,
we are in the localized regime. The mean free path and the localization length
both depend strongly on the energy interval under consideration and on the
disorder strength, see the discussion in Refs.~\cite{sheng,2D}.

In our multidimensional arrangement, a double super-radiant transition again
occurs as in $1d$, which we do not discuss in detail here. In this Section we
will focus on the dependence of the maximum conductance on the coupling
strength to the leads. The dimensionless conductance $G$, which is proportional
to the total transmission, can be computed using the Landauer
formula~\cite{Beenakker,epl},
\begin{equation}
G(E)=\sum_{a=1}^{M} \sum_{b=M+1}^{2M}  |Z^{ab}(E)|^2 \,.
\label{G}
\end{equation}
Here $Z^{ab}(E)$ is the transmission amplitude between channels $a$ and $b$,
see Eq.~(\ref{T2}), that can be computed from the effective Hamiltonian.

\begin{figure}[h!]
\vspace{0cm}
\includegraphics[width=7cm,angle=-90]{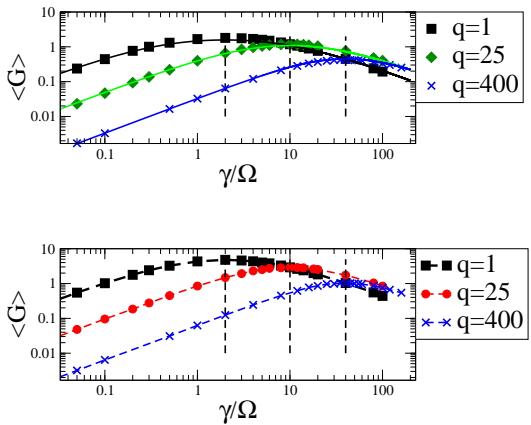}
\caption{ (Color online) The average dimensionless conductance as a function of
the coupling strength $\gamma / \Omega$. In the upper panel we consider the
quasi-$1d$ case of $10 \times 100$ sites in the diffusive regime, with disorder
$W/\Omega= \sqrt{3/4}$, so that $L/l \approx 3$. The analytical results of
random matrix theory ~\cite{Beenakker} (solid lines) are compared with our
numerical results (symbols). In each case, an overall multiplicative factor in
the analytical expression~\cite{Beenakker} has been obtained by fitting to the
numerical data. In the lower panel, we consider the $2d$ case with $30 \times
30$ sites and $W/\Omega=2$, so that $L/l \approx 4$. Here symbols refer to our
numerical results and the dashed lines simply connect the symbols (random
matrix results are not available for asymmetric $2d$ systems in the diffusive
regime). Vertical dashed lines indicate the critical values $\gamma/ \Omega= 2
\sqrt{q}$, where the maximum conductance is expected for each asymmetry value
$q$. All numerical results are obtained by averaging over $200$ realizations of
disorder and over $100$ different energies in the interval $- 0.1<E <0.1$
around the center of the energy band. } \label{DiffAsym}
\end{figure}

We will not consider here the case of a small perturbation to an integrable
Hamiltonian, which displays very system-specific behavior,
and start by analyzing the case of moderate disorder. In Fig.~\ref{DiffAsym},
we illustrate the behavior of the average conductance in the diffusive regime
for a quasi-$1d$ lattice of $M\times L=10 \times 100$ sites with
$W/\Omega=\sqrt{3/4}$, upper panel, and for a square $2d$ lattice of $30 \times
30$ sites with $W/\Omega=2$, lower panel. In both cases, the maximum
conductance is obtained at the critical value of the coupling given by
Eq.~(\ref{gcr3}). In the upper panel, we compare our numerical results with
analytical expressions obtained in the context of random matrix
theory~\cite{Beenakker}. The numerically computed left and right transmission
coefficients as functions of $\gamma / \Omega$ were used to evaluate the
analytical expression for the average conductance given in
Ref.~\cite{Beenakker}. In the upper panel of Fig.~\ref{DiffAsym}, the numerical
results (symbols) are in good agreement with the analytical results (solid
curves) for a quasi-$1d$ system in the diffusive regime. To the best of our
knowledge, there are no known analytical results for the $2d$ case in the
diffusive regime with asymmetric coupling (lower panel).

At the critical value of $\gamma/ \Omega$ given by Eq.~(\ref{gcr3}), each
conductance curve for the asymmetric case intersects the conductance curve for
the symmetric case ($q=1$). This is similar to what we have observed in the
$1d$ case (see Fig.~\ref{Asym1D}), and is also consistent with the results
obtained in the context of random matrix theory. As discussed in
Sec.~\ref{sec_RMT}, the maximum conductance is achieved when $\tau_1 = \tau_2$.
In this case a system with asymmetric coupling behaves as a system with
symmetric coupling.

\begin{figure}[h!]
\vspace{0cm}
\includegraphics[width=7cm,angle=-90]{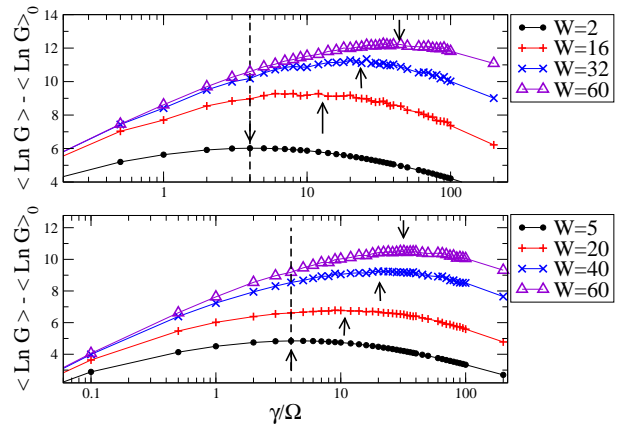}
\caption{ (Color online) Average logarithm of the conductance as a function of
the coupling strength $\gamma/ \Omega$ for the multidimensional case, with
$\Omega=1$, asymmetry parameter $q=4$, and different disorder strengths $W$.
For the sake of clarity, we plot $\langle \ln G \rangle-\langle \ln G
\rangle_0$ so that all data start from the same point, where $\langle \ln G
\rangle_0$ is defined as in Fig.~\ref{surenW}. In the upper panel, we show the
$2d$ case with $20 \times 20$ sites, while in the lower panel we show the $3d$
case, with $8 \times 8 \times 8$ sites. The dashed vertical line shows the
critical value of the coupling, Eq.~(\ref{gcr3}), at which we have maximum
conductance for moderate disorder. The arrows indicate $\gamma/\Omega=0.76 ND
/\Omega$ for the $2d$ case, upper panel, and $ 0.56 ND/\Omega $ for the $3d$
case, lower panel. The numerical factors $0.76$ and $0.56$ have been found from
fitting. All data have been obtained by averaging the conductance over $10^4$
ensemble realizations at the center of the energy band, $E=0$. } \label{MattW}
\end{figure}

In the regime of very strong disorder, the critical value at which we have
maximum conductance becomes dependent on the mean level spacing, $D$. Similarly
to the $1d$ case we have
\begin{equation}
\gamma_{\rm cr} \propto ND  \,.
\label{RMg}
\end{equation}
Since  in this regime $D \approx W/N$, we find that the critical coupling for
the maximum conductance is proportional to the disorder strength $W$, see
Fig.~\ref{MattW}.

For strong disorder, we enter the localized transport regime, where the
transmission is log-normally distributed~\cite{Beenakker}, so that, similarly
to Fig.~\ref{surenW}, it is more convenient to consider the average of $\ln G$ rather than
the average of $G$. In Fig.~\ref{MattW} we plot the average logarithm of the
conductance as a function of $\gamma/ \Omega$ in $2d$, upper panel, and in
$3d$, lower panel. As indicated by the arrows, the critical value of  $\gamma/
\Omega$ at which the conductance is maximized is in both cases proportional to
the mean level spacing, which in turn is proportional (for strong disorder) to
the disorder strength. For moderate disorder in either $2d$ or $3d$, see e.g.
$W=2$ in the upper panel and $W=5$ in the lower panel, the critical value of
$\gamma/ \Omega$ for which we have the maximum conductance is again given by
Eq.~(\ref{gcr3}). A detailed comparison between random matrix results and the
Anderson model will be presented elsewhere.

\section{Conclusion}

We have analyzed coherent quantum transport through a finite sequence of
potential barriers with asymmetric coupling to the external environment. As the
coupling to the environment is varied, transport properties are greatly
affected. In a previous paper~\cite{Kaplan}, the super-radiant transition that
occurs in this paradigmatic model, at a critical value of the coupling, was
studied for the case of symmetric coupling to environment. Here, with the aid
of the effective non-Hermitian Hamiltonian, we show that for asymmetric
coupling a double super-radiant transition occurs, as compared with a single
transition in the symmetric case.

The super-radiant transitions have important consequences for the observable
resonance structure. In particular, the number of resonances decreases by one
after each transition. Focusing on the behavior of transmission near the
energy band center, see Fig.~\ref{Tmax}, we demonstrate that a sharp change in
the structure of the resonance occurs in correspondence with the two
super-radiant transitions. This change can be characterized by the behavior of
the transmission maxima and minima, which we describe analytically as functions
of the coupling to the environment and of the asymmetry of this coupling. As
far as we know, these features of the structure of resonances as a function of
the coupling strength to the leads have not previously been reported in the
literature.

Maximum transmission through the system is reached at a coupling $\gamma /
\Omega = 2 \sqrt{q}$, where $q$ is the asymmetry parameter. This coupling is equal to the
geometric mean of the coupling strengths associated with the two transitions. We show
that this result does not follow from random matrix theory, as usually assumed
in statistical theories of quantum transport. Moreover, for  very strong
disorder we show that the coupling at which transmission is maximized is
proportional to the disorder strength. We also find that the latter results
remain valid in higher dimensions. Specifically, we analyze the average
conductance as a function of the degree of opening, of the asymmetry, and of
the strength of disorder in the multidimensional cases: quasi-$1d$, $2d$, and
$3d$. In the quasi-$1d$ case, we compare our results with analytical
expressions obtained in the context of random matrix theory. We demonstrate the
validity of our results in $2d$ and $3d$, in  both diffusive and localized
transport regimes, where, to the best of our knowledge, no analytical results
as a function of the asymmetry of the opening are available in the literature.

The results presented here are based on an approach originally formulated in
the framework of nuclear reaction theory. Now we understand that they reflect
general properties of quantum signal transmission. Therefore they might
be of relevance for numerous applications. The sequence of potential barriers
is a paradigmatic model for coherent quantum transport. A better understanding
of this transport regime is essential for the development of information
technology using different nanoscale systems with complex geometry, including
quantum dots, photonic crystals, and molecular wires.

\section*{Acknowledgments}

We acknowledge very useful discussions with  F.~M.~Izrailev.
This work was supported in part by the NSF under Grants
PHY-0545390 and PHY-0758099. S.~Sorathia
acknowledges financial support from the Leverhulme Trust.
G.~L.~C. acknowledges financial support from E.U.L.O.
V.~Z. is thankful to V. V. Sokolov for constructive discussions.

\appendix \section{Analytical Derivation of the Transmission in the Ordered Case}

To derive closed analytical expressions for the sums (13), it is convenient to
double them, making the phase $\varphi_{n}=n\pi/(N+1)$ run over the whole
circle. Then
\begin{equation}
P_{\pm} (\epsilon) = \frac{1}{2 \Omega}  \frac{1}{2 N + 2}
\sum_{n = 0}^{2 N + 1} (\pm 1)^n \frac{\sin^2(\varphi_n)}{\epsilon - \cos(\varphi_n)}\,,
                                  \label{A1}
\end{equation}
where $\epsilon=E/2\Omega$. Now the sum over $n$ spans all the roots $z_n =
\exp(i \varphi_n)$ of the equation $z^{2 N + 2} - 1 = 0$, so that
\begin{equation}
P_{\pm} (\epsilon) = \frac{1}{4 \Omega}  \frac{1}{2 N + 2}  \sum_{n = 0}^{2 N + 1}
(\pm 1)^n \frac{(z_n^2 - 1)^2}{z_n (z_n^2 - 2 \epsilon z_n + 1)}\,.
                                   \label{A2}
\end{equation}
Splitting the summand into simple fractions and using the identities
\begin{equation}
\frac{1}{2 N + 2} \sum_{n = 0}^{2 N + 1} (\pm 1)^n = \frac{(1 \pm 1)}{2}\,,
                                    \label{A3}
\end{equation}
\begin{equation}
\frac{1}{2 N + 2} \sum_{n = 0}^{2 N + 1} (\pm 1)^n z_n = \frac{1}{2 N + 2}
\sum_{n = 0}^{2 N + 1} (\pm 1)^n  \frac{1}{z_n} = 0\,,
                                    \label{A4}
\end{equation}
\begin{equation}
\frac{1}{2 N + 2} \sum_{n = 0}^{2 N + 1} (\pm 1)^n \frac{1}{z_n - z} = -
\frac{1}{2} \left( \frac{z^N}{z^{N + 1} - 1} \pm \frac{z^N}{z^{N + 1} + 1}
\right)\,,                          \label{A5}
\end{equation}
we can write down the original sums in closed form,
\begin{equation}
P_+ (\epsilon) = \frac{1}{2\Omega} \left\{ \epsilon - \frac{2 (\epsilon^2 - 1)}{z_{+} - z_{-}}
\left[ \frac{z_{+}^{2 N + 2}}{z_{+}^{2 N + 2} - 1} - \frac{z_{-}^{2 N + 2}}{z_{-}^{2 N + 2} - 1}
\right] \right\}\,,                   \label{A6}
\end{equation}
\begin{equation}
P_- (\epsilon) = - \frac{1}{\Omega}  \frac{(\epsilon^2 - 1)}{z_{+} - z_{-}}
\left\{  \frac{z_{+}^{N + 1}}{z_{+}^{2 N + 2} - 1} - \frac{z_{-}^{N + 1}}{z_{-}^{2 N
+ 2} - 1} \right\}\,,               \label{A7}
\end{equation}
where $z_{\pm}=\epsilon \pm \sqrt{\epsilon^2-1}$ are the roots of the quadratic equation $z^2 - 2 \epsilon z + 1
= 0$, and $z_{+}z_{-}=1$. In the asymptotics of large $N$ and energy outside
the Bloch band, $|\epsilon|>1$, we have either $|z_{+}^N| \gg 1 \gg |z_{-}^N|$
or $|z_{-}^N| \gg 1 \gg |z_{+}^N|$, for $\epsilon>1$ or $\epsilon <-1$ respectively, while
inside the band we have $z_{\pm}= \exp(\pm i\beta)$. This leads to Eqs.~(\ref{pplusoutside}) - (\ref{exN})
in the main text.

To calculate the transmission amplitude (\ref{a}) or transmission probability (\ref{d}), we also need
\begin{equation}
P_{+}^{2}(\epsilon)-P_{-}^{2}(\epsilon)=\frac{1}{4\Omega^{2}}\,
\frac{\sin[(N-1)\beta]}{\sin[(N+1)\beta]}\,,           \label{44}
\end{equation}
valid for all energies inside the band, $|\epsilon|<1$. For $N\gg 1$, we
can consider averaging over a small energy interval,
similarly to Eq.~(\ref{tavgmiddle}), where $\epsilon=\cos\beta$ is constant while the trigonometric
functions depending on $N\beta$ change from $-1$ to $+1$. This can be done formally
with the substitution $\beta\rightarrow\beta+\xi/N$, where $\xi$ is of
order one. Then the average transmission $\overline{T} \equiv \overline{T^{12}}=\overline{|Z^{12}|^2}$ is given by the integral
\begin{equation}
\overline{T^{12}}=4\eta_{1}\eta_{2}\int
\frac{d(N\beta)}{\pi}\,\frac{1}{\Phi(N\beta)}\,,                \label{45}
\end{equation}
where $\eta_{1,2}=\gamma_{1,2}/2\Omega$ and
\begin{equation}
\Phi(y)=A^{2}\sin^{2}y+B^{2}\cos^{2}y+2C\sin y\,\cos y \,,    \label{46}
\end{equation}
\begin{equation}
A^{2}=(\eta_{1}+\eta_{2})^{2}+(1-\eta_{1}\eta_{2})^{2}\cos^{2}\beta \,, \label{47}
\end{equation}
\begin{equation}
B^{2}=(1+\eta_{1}\eta_{2})^{2}\sin^{2}\beta \,,                \label{48}
\end{equation}
\begin{equation}
C=(1-\eta_{1}^{2}\eta_{2}^{2})\cos\beta\,\sin\beta \,.       \label{49}
\end{equation}
After changing the integration variable to $x=\tan(N\beta)$, the integral becomes
\begin{equation}
I(\beta)=\frac{1}{\pi}\,\int_{-\infty}^{\infty}\frac{dx}{A^{2}x^{2}
+2Cx+B^{2}}=\,\frac{1}{\sqrt{A^{2}B^{2}-C^{2}}} \,.            \label{50}
\end{equation}
This leads to the final result for the average transmission at
any energy inside the band,
\begin{equation}
\overline{T^{12}}=\,\frac{4\eta_{1}\eta_{2}\sin\beta}
{(\eta_{1}+\eta_{2})(1+\eta_{1}\eta_{2})} \,,                \label{51}
\end{equation}
where $\sin\beta=\sqrt{1-\epsilon^{2}}$,
to be compared to Eq.~(\ref{tavgmiddle}). By means of slightly more complicated integrals, we derive
[here $\cos(2\beta)=2\epsilon^{2}-1$]
\begin{equation}
\overline{T^{11}}=4\eta_{1}^{2}\,\frac{\eta_{1}(1+\eta_{1}\eta_{2}+\eta_{2}^{2})
+\eta_{2}\cos(2\beta)+(1+\eta_{2}^{2})\sin\beta}{(\eta_{1}+\eta_{2})(1+\eta_{1}
\eta_{2})(1+\eta_{1}^{2}+2\eta_{1}\sin\beta)} \,,             \label{52}
\end{equation}
and similarly for $\overline{T^{22}}$. This, along with Eq.~(\ref{a}), allows one to check 
the unitarity condition,
\begin{equation}
T^{11}+T^{12}=-2\,{\rm Im}\,Z^{11} \,,             \label{53}
\end{equation}
which remains valid after averaging, where both sides are equal to
\begin{equation}
\frac{4\eta_{1}(\eta_{1}+\sin\beta)}{1+\eta_{1}^{2}+2\eta_{1}\sin\beta} \,.
                                                         \label{54}
\end{equation}


\end{document}